# Two-dimensional semiconducting nanostructures based on single graphene sheets with lines of adsorbed hydrogen atoms

*Leonid A Chernozatonskii*[*,1], *Pavel B Sorokin*[1,2,3], *Jochen W Brüning*[4]

[1]Emanuel Institute of Biochemical Physics, Russian Academy od Sciences, 4 Kosigina st., Moscow, 119334, Russia

[2]Siberian Federal University, 79 Svobodniy av., Krasnoyarsk, 660041 Russia and

[3]Kirensky Institute of Physics, Russian Academy of Sciences, Academgorodok, Krasnoyarsk, 660036 Russia

[4]Institute of Mathematics, Humboldt University of Berlin, Berlin, 12489 Germany

It is shown that lines of adsorbed hydrogen pair atoms divide the graphene sheet into strips and form hydrogen-based superlattice structures (2HG-SL). We show that the forming of 2HG-SL drastically changes the electronic properties of graphene from semimetal to semiconductor. The electronic spectra of "zigzag" ($n$,0) 2HG-SL is similar to that of ($n$,0) carbon nanotubes and have a similar oscillation of band gap with number $n$, but with non-zero minimal values. The composite dual-periodic ($n$,0)+($m$,0) 2HG-SLs of "zigzag" strips are analyzed, with the conclusion that they may be treated as quasi-two-dimensional heterostructures. We also suggest an experimental way of fabricating hydrogen superlattices.

The miniaturization of electronic devices has always been a driving force in electronic industry. The exciting discovery by Geim and Novoselov [1, 2] of the existence of single graphite sheets has opened new possibilities in fabricating of nanoelectronic devices based on graphene. It is outstanding due to its extraordinary electronic properties: the energy spectrum contains the so-called Dirac points, i.e., K points of the connection of band cones, to which the Fermi surface degenerates. It leads to number of attractive properties of graphene, i.e. unusual quantum Hall effect, ballistic transport [1] and etc. In the last few years a great deal of works has been dedicated to the investigation of the properties of graphene [3].

One of the many interesting properties of Dirac electrons in graphene is the drastic changes of the conductivity of graphene-based structures with the confinement of electrons. Until now, two possibilities for the realization of this effect have been proposed: carbon nanotubes (CNT) [7] (periodic boundary conditions for the wave-vector of the electron) and graphene ribbons prepared by using electron lithography methods [8, 9] (zero boundary conditions). In these structures, the band gap oscillates depending on the size, with either zero and non-zero minimal values, for carbon nanotubes [7] or graphene ribbons [10], respectively.

---

[*] cherno@sky.chph.ras.ru

In this Letter, we propose another possibility of electron confinement using chemical adsorption of H atom pairs on the surface of graphene. Hydrogen lines split the graphene sheet into ribbons (electronic waveguides) with same or different electronic properties, forming two-dimensional (2D) superlattices - see Fig. 1.

Until now 2D superlattices with one-dimensional channels was fabricated by various methods in different materials, e.g., in alkali-metal-doped conjugated polymers [4], on stepped Au (111) surfaces [5], and by modulating the electrostatic potential in 2D electron gas by surface acoustic waves in GaAs-$Al_xGa_{1-x}As$ [6], but all superlattices prepared earlier have ~ 1 micron period length. Moreover, they were not pure two-dimensional objects. For instance, the "layer by layer" technique can fabricate the heterostructures with 2D electrons localized near the interface in 3D atomic framework whereas proposed 2D superlattices based on graphene – on the object with single-atom layer thickness.

We show that free-standing graphene ruled by lines of covalently bonded pairs of hydrogen atoms has electronic properties similar to carbon nanotubes. The "zigzag" superlattice structures are semiconductors. Changing of width strip, i.e. period of such superlattice, leads to changing of the energy gap width of the 2H-line graphene-based superlattice (2HG-SL). Thus we can obtain set of 2D-semiconductors with different properties. The present study is based upon the recent experiments with free-standing graphene sheets [11] and hydrogen dimer adsorption on graphite during annealing [12]. In another paper, it has been concluded that the barrier of hydrogen chemisorption should decrease with the elongation of hydrogen line [13].

The preparation of carbon nanotubes with adjusted electronic structure is still unsolved problem. At the end of the paper we propose the simple method of preparation 2HG-SLs which should allow making such superlattice with prescribed electronic properties. We believe that the growing of nanotechnology toolkit will open the opportunity of preparation of such attractive graphene-based nanostructures.

Our calculations were performed using density functional theory [14, 15] within the local density approximation for the exchange-correlation functional [16], employing norm-conserving Troullier-Martins pseudopotentials [17] in the Kleinman-Bylander factorized form [18]. Finite- range numerical pseudoatomic wave functions were used as an atomic-orbital basis set. Slabs were treated in a supercell scheme allowing enough empty space between them to make intermolecular interactions negligible. The geometry of the structures was optimized until residual forces became less than 0.04 eV/Å. The real-space mesh cutoff was set to at least 175 Ry. The Monkhorst-Pack [19] special *k*-point scheme was used with 0.08 Å$^{-1}$ *k*-point spacing.

We used the SIESTA package [20, 21] in all calculations. All the values given above were carefully tested and found optimal.

We have investigated the dependence of electronic properties of 2HG-SLs by considering "zigzag" ($n$,0) 2HG-SLs as well as dual ($n$,0)+($m$,0) 2HG-SLs (see Fig.1a). Here, 2H lines divide a graphene sheet into zigzag ($n$,0) strips (this notation is directly related to the notation commonly adopted for CNTs [7]). We have also checked the properties of free-standing ($n$,$n$) 2HG-SLs, which turn out to be semimetallic.

First, we consider ($n$,0) 2HG-SLs. The geometric scheme of these structures is shown on Fig. 1b. Hydrogen atoms, shown in blue, are covalently bound to C atoms, shown in cyan, forming lines perpendicular to the ($n$,0) direction in graphene. A similar scheme has been used for the investigation of hydrogenated ($n$,0) CNTs [22, 23]. The H-atoms form local $sp^3$-hybridization between hydrogen and carbon atoms, which causes a local geometrical distortion of the graphene sheet, as if forming diamond-like lines.

In Fig. 1(b-d), the electronic band structure of the (8,0) 2HG-SL (Fig. 1b) is compared with that of a (8,0) ribbon (in the conventional notation, 17-AGNR (armchair graphene nanoribbon) [24, 10]) (Fig. 1c) and to the spectrum of a (8,0) CNT (Fig. 1d). There is a close resemblance between all spectrums, however in the case of superlattice, the band gap is roughly twice as large as that of CNT ($E_{CNT}$ = 0.22 eV, $E_{2HG-SL}$=0.39 eV). We believe that this behavior results from confinement of electrons by presence of $sp^3$ - carbon-hydrogen lines (we don't compare the gap widths of the superlattice and the ribbon as the latter has a different behavior of the band gap [24, 10]).

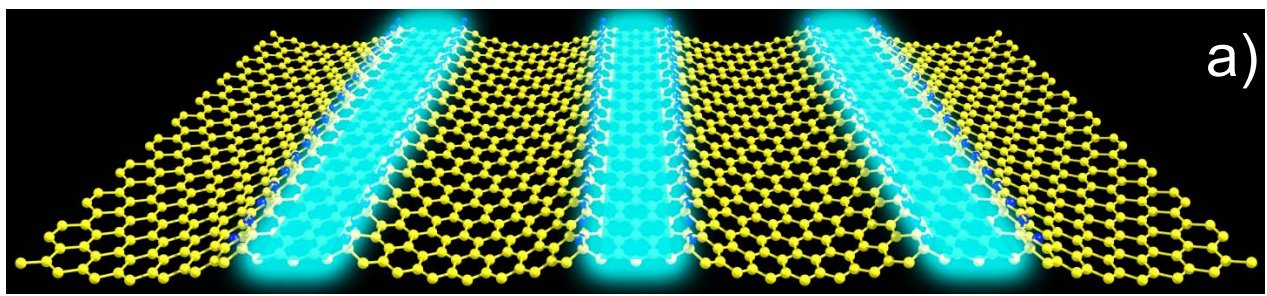

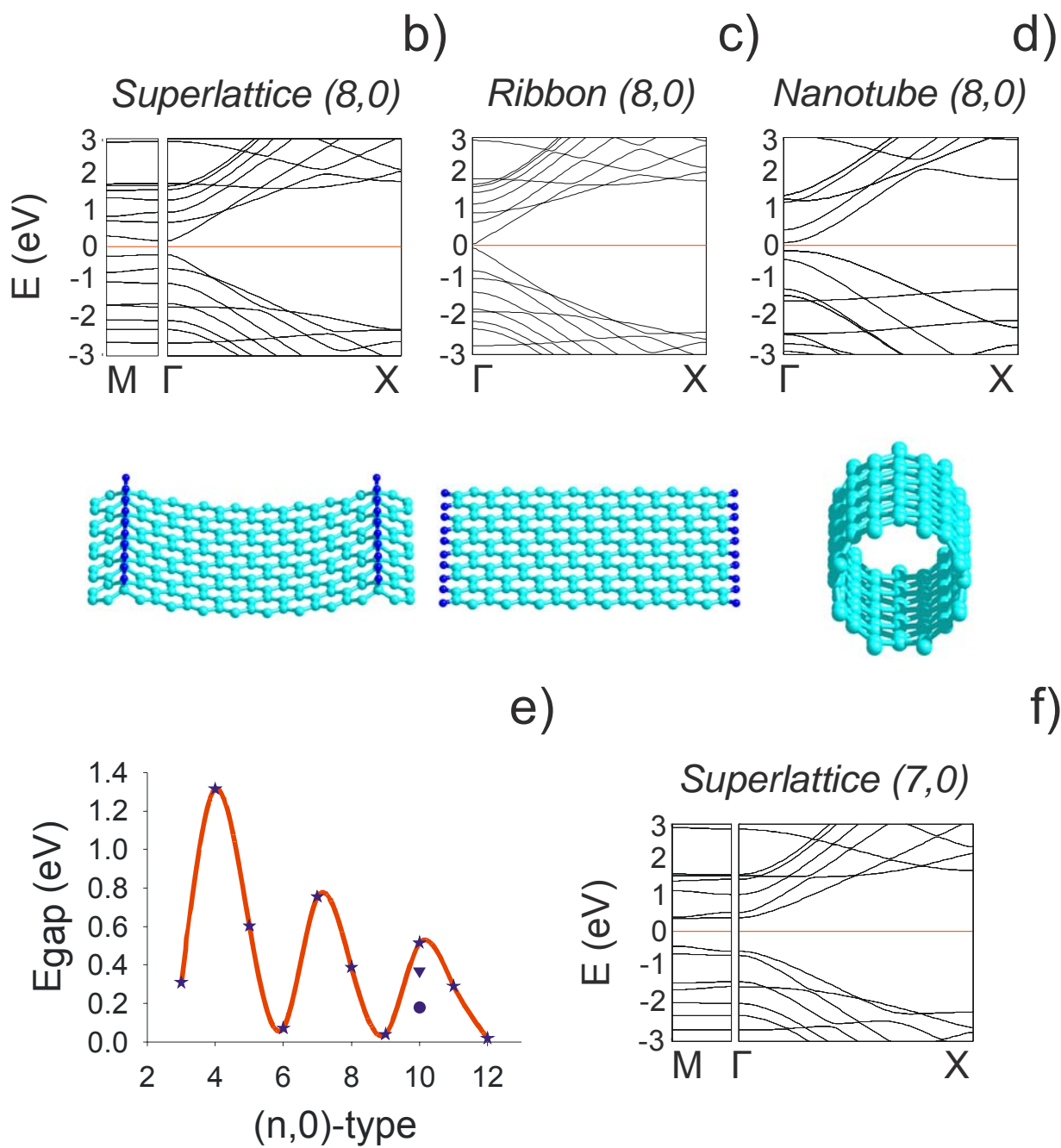

Fig. 1. Geometry and electronic structure of 2H-line superlattice models: a) (3,0)+(7,0) 2HG-SL scheme, where (3,0) strips (highlighted) are electronic waveguides. Band structures and corresponding views of b) (8,0) 2HG-SL, c) graphene ribbon (17-AGNR in conventional notation [24]), and d) (8,0) carbon nanotubes, (carbon atoms are cyan, hydrogen atoms are blue); e) the variation of band gaps of ($n$,0)- 2HG-SLs as with index $n$ (the triangle and the circle correspond to band gap width of (3,0)+(7,0) and (4,0)+(6,0) 2HG-SLs, respectively); f) the band structure of (7,0) 2HG-SL. The Fermi level in all spectra ($E$ = 0) is marked by the horizontal line.

Calculations have been performed for a set of ($n$,0) 2HG-SLs with $n = 3 - 12$. We have found that band gaps oscillate with increasing superlattice width and vanish in the infinite limit of pure, semimetallic graphene (see Fig. 1e). We want to mention that the oscillation of band gap of ($n$,0) 2HG-SLs is more similar to carbon zigzag ($n$,0) nanotube gaps oscillation [7] than to the behavior graphene ribbon gaps [24]. In the 3$n$ case, small but nonzero gaps are obtained. We suppose that the existence of nonzero gaps in the 3$n$ case is connected with specific C-H bond edge effects, similarly to graphene ribbons [10]. The (3$n$ + 1) 2HG-SLs with maximal band gaps have a direct gap at the M point of the spectrum whereas other 2HG-SLs have one in Γ point - compare the electronic spectra of (7,0) and (8,0) superlattices at Fig. 1 (f and b).

The discussion of the origin of 3$n$ dependence is important and interesting. In previous work [25] where the 2HG-SL layer with substrate system has been investigated, van der Waals interactions cause flattening of graphene. Therefore such a system can be roughly treated like a sequence of graphene ribbons with finite barriers between neighbors formed by $sp^3$-hybridized carbon atoms with strained bonds to neighboring $sp^2$-atoms. In the case of graphene ribbons, the oscillation of the band gap has a 3$n$+2 behavior, and for the graphene-substrate system we have obtained the same result. But in the present structures, we have a completely different situation (see Fig. 1e). We believe that 3$n$ dependence originate from curvature of a 2HG-SL due to the absence of substrate. The 3$n$ and 3$n$ + 2 behaviors of small-period free-standing 2HG-SLs correlate with carbon nanotubes with small diameter and graphene ribbons, respectively. Thus the 2HG-SL can be considered an intermediate state between these structures. The role of curvature in the electronic properties of carbon nanotubes is well known fact [7], e.g., it has been found in Ref. [26] the knowing only the curvature is enough to predict carbon nanotube properties. It may be assumed that the dependence of 2HG-SL band gap can be switched from 3$n$ to 3$n$ + 2 modes by longitudinally straining the superlattice or using different substrates which influence the curvature of a 2HG-SL.

A common property of all considered modified "2H-line" graphene 2HG-SLs is the anisotropy of bands along different directions (see, e.g., the band behavior along M-Γ and Γ-X directions at Fig. 1b and Fig. 1f). The anisotropy of the 2HG-SL bands (Fig. 1b and Fig. 1f) is similar to that commonly found in superlattices of different material layers, like in $Ga_xAl_{1-x}As$ [27], but with low semiconductor gaps < 1 eV (infrared region).

We have also studied composite 2HG-SLs consisting of 2H-bordered strips with different widths (see Fig. 2). The electronic spectrum E($k$) of such structures along graphitic strips behaves similar to the ($n$,0)-2HG-SL spectra and isn't shown. The location of conduction band bottom and valence band top is determined by the strip with the larger band gap width. For instance, the (4,0)+(6,0) structure has a direct band gap at the M point due to the presence of the (4,0) strip, whereas the (3,0)+(7,0) structure has a direct gap at the Γ point since the (7,0) strip has a direct gap at the origin. Here, the band gap width is dictated by the strip with the narrower gap:

$E_{gap}^{(4,0)+(6,0)} = 0.18$ eV and $E_{gap}^{(5,0)+(6,0)} = 0.18$ eV;

$E_{gap}^{(6,0)} = 0.07$ eV $< E_{gap}^{(5,0)} = 0.60$ eV $< E_{gap}^{(4,0)} = 1.32$ eV;

$E_{gap}^{(3,0)+(7,0)} = 0.37$ eV,

$E_{gap}^{(3,0)} = 0.31$ eV $< E_{gap}^{(7,0)} = 0.76$ eV.

The difference of band gaps of connected strips also forms regions with different conduction electron density, like in 1D superlattices described earlier (e.g. carbon nanotubes with alternate adsorbed hydrogen [22] or BN-C nanotubes connection [28]). The geometrical scheme, highest occupied molecular orbital (HOMO) and lowest unoccupied molecular orbital (LUMO) distribution are shown in Fig. 2 for (4,0)+(6,0) and (3,0)+(7,0) structures. All orbitals are shown for the Γ point. In the case of the (4,0)+(6,0) 2HG-SL, corresponding electrons are localized near the (6,0)-strip, which, being rolled, forms the semimetallic carbon nanotube (6,0). The more "dielectric" (4,0) strip is nearly empty. Thus the (4,0)+(6,0) composite 2HG-SL may be treated as a heterostructure. In Fig. 2a, the bottom, the schematic band diagram of such structure is shown. We have used the technique of energy band diagram description proposed in Ref. [22], where the band offset was directly obtained from the local density of states. The "dielectric" (4,0) strip has low but nonzero values of local density of states in the "gap" region due to very small strip width and tunneling effect of electrons through the diamond-like barrier. The tunneling effect should be suppressed in the 2HG-SL with bigger width of the "dielectric" strip.

The (3,0)+(7,0) structure (see Fig. 1a and Fig. 2b) has the same period of 10 hexagons, however it differs from the (4,0)+(6,0) 2HG-SL only in the single-hexagon shift of 2H-lines. In this case, the molecular orbital distribution is completely different. Now the (3,0) strip is filled, whereas the (7,0) strip is empty.

This effect demonstrates that an electron waveguide, a quantum nanometer wire with a "single atom thickness", can be created by decorated a "quasi-metallic" (3n,0) strip on graphene by two "dielectric" (3n + 1,0) or (3n + 2,0) strips by its sides. Thus, it can be possible to construct integrated circuits using various combinations of connected waveguides on just a single graphene sheet.

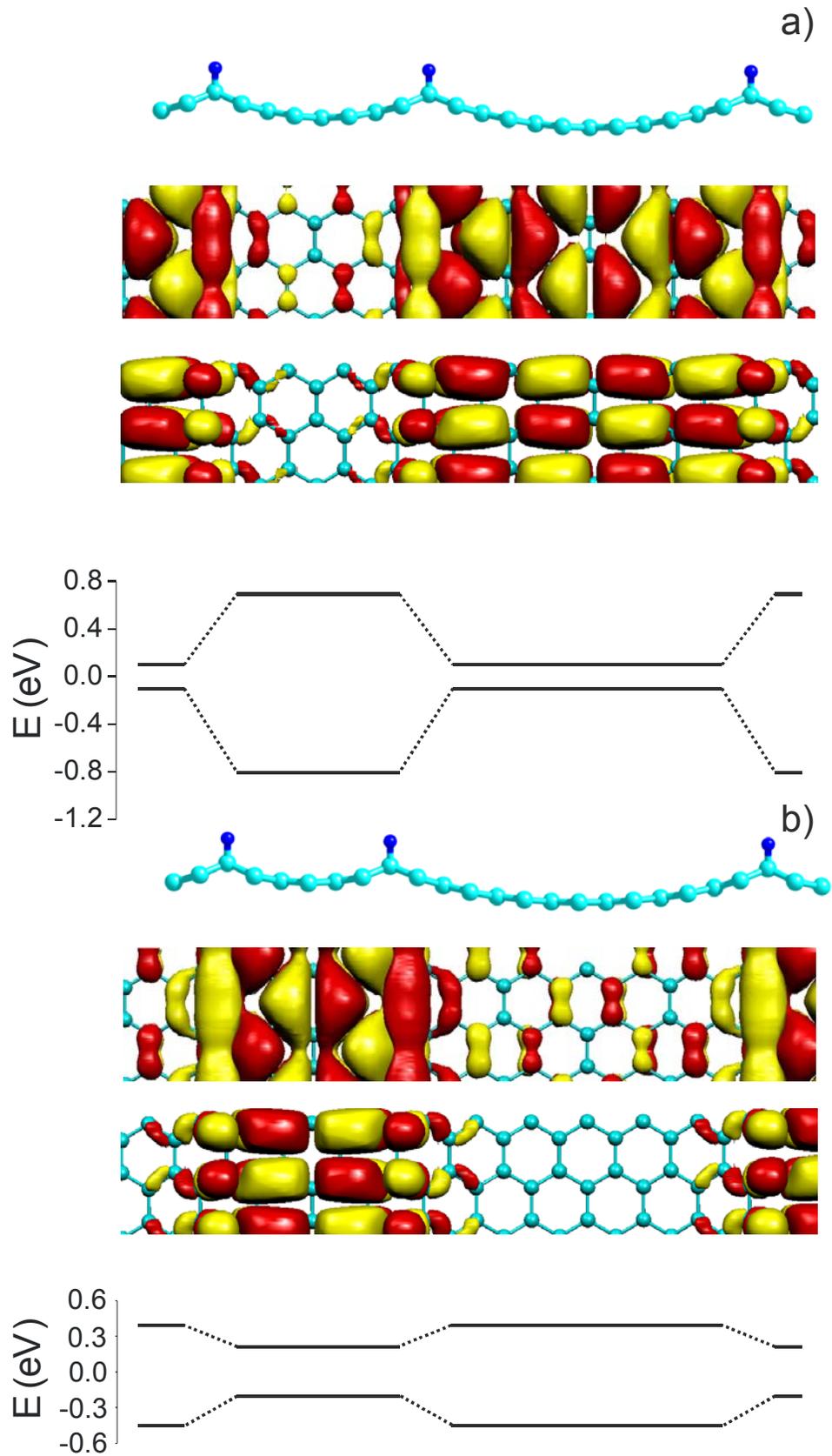

**Fig. 2.** The geometrical structure, HOMO and LUMO orbitals at the Γ point (isovalue 0.02) and a schematic illustration of the energy band diagram of a) (4,0)+(6,0) and b) (3,0)+(7,0) 2HG-SLs.

We propose a possible method of 2HG-SL fabrication. It has been shown both theoretically [29] and experimentally [30] that the probability of hydrogen chemisorption increases with increasing local curvature of the $sp^2$ carbon network.

Then, the bending of a graphene sheet should increase the probability of hydrogen adsorption on its surface. The curvature of the sheet can be changed using a step on the substrate. The formed step of graphite substrate (or of some crystalline substrate) will modify the superposed graphene sheet as shown in Fig. 3a. The curvature of graphene in region 2 is positive; therefore this region can be treated like a fragment of a carbon nanotube. Other regions are unfavorable for adsorption: regions 1 and 4 have zero curvature and a low chemical activity, whereas in region 3 it has a negative curvature that hinders the formation of $sp^3$- hybridization [29]. Thus, a hydrogen line should only form in the second region during the experiment. Longitudinal shifting of graphene should flatten this region and bend another part of the sheet, thus making it possible to form periodically arranged hydrogen lines step by step.

To describe such structures, we have used molecular mechanic method [31], which gives good qualitative description of geometrical and elastic properties of large carbon structures (containing more than one thousand C-atoms) [32].

When a single graphite layer step is used, the curvature radius in region 2 is equal to 18 Å. This value corresponds to the (46,0) zigzag nanotube. A double footstep, shown in Fig. 3b, increase the graphene curvature a little further – the radius of curvature decreases to 15 Å. This value corresponds to the (38,0) zigzag tube. The longitudinal compression of the upper sheet (Fig. 3c) leads to a further increase of the curvature even in the case of one graphite layer step. A strain value of 2% result in a significant decrease of region 2 curvature radius up to 8 Å, which is close to the radius of the (20,0) tube.

As-fabricated superlattices can be transferred over preliminarily prepared grooves, e.g., on the surface of $SiO_2$ [11]. Thus, free-standing graphene-based superlattice structures with different strip widths for use as tunneling elements in nanoelectronics can be prepared.

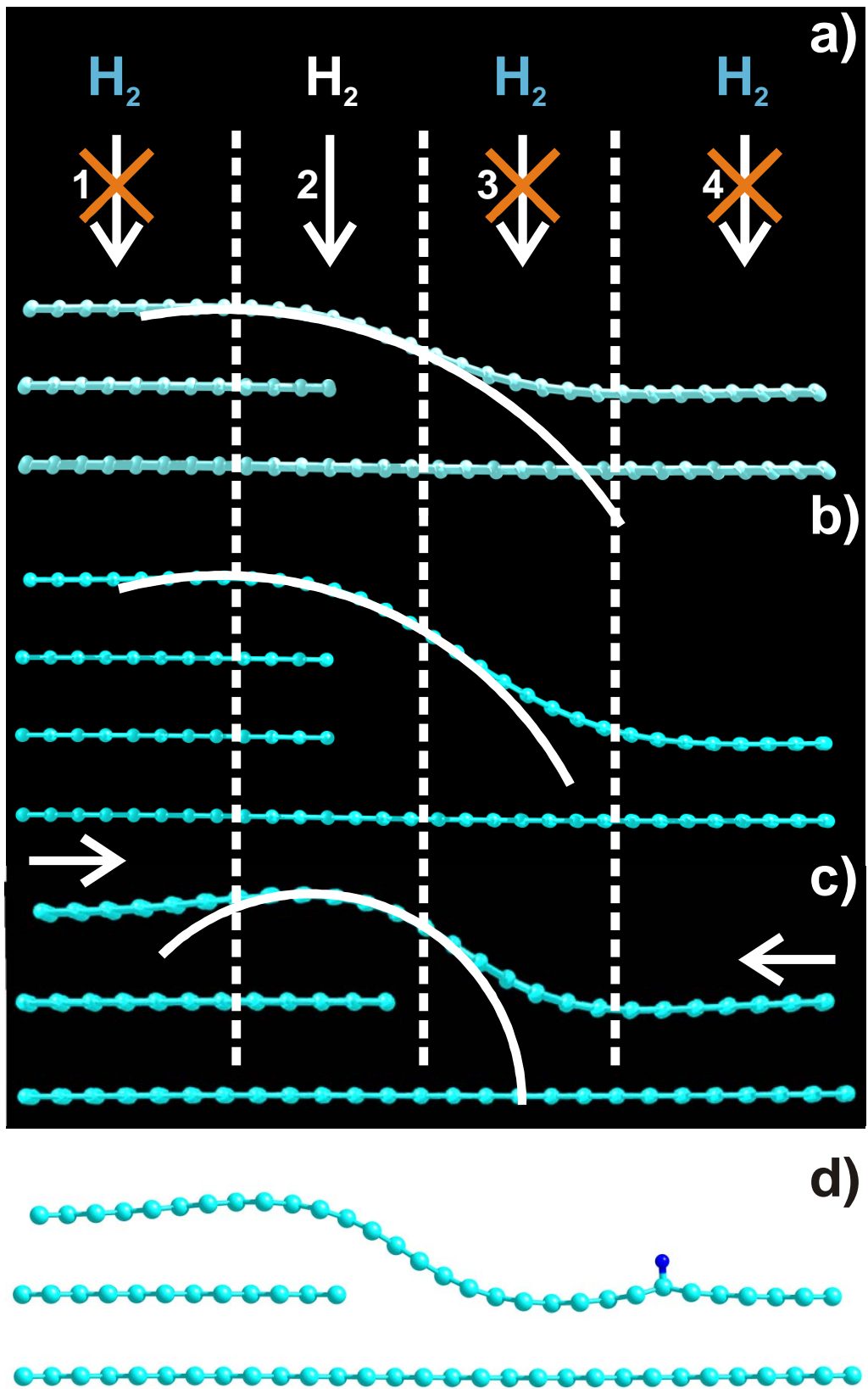

Fig. 3. The scheme of fabrication of graphene superlattices: regions 1 and 4 are flat, while 2 and 3 have positive and negative curvature, respectively. The possibilities of further increasing of graphene curvature from a) the one step case up to b) double step case or c) longitudinal strain using. d) A shifted sheet of graphene with a single adsorbed hydrogen line, ready for the next 2H adsorption procedure.

Finally, we discuss the very important question of stability of such structures. DFT calculations show that the energy penalty (the difference of initial and final energies) for a hydrogen atom to leave the line and move away to another carbon atom is rather high, ~ 1.9 eV – not to mention the chemical reaction barrier. It is 1.5 times as much as the corresponding difference for H in a single adsorbed hydrogen dimer (1.17 eV [12]). This is a clear proof of high stability of considered 2HG-SL structures.

It has been shown that covalently bonded pairs of hydrogen atoms on the uncombined graphene may form strips with nanometer periods, and significantly change the semimetallic spectrum of pure graphene: 2HG-($n$,0) superlattices are semiconductors with a gap that depends on their period. The proposed 2HG-SLs possess electronic properties that are intermediate between that of carbon nanotubes and graphene ribbons. They have a 3$n$ band gap oscillatory behavior, like carbon nanotubes, but non-zero gaps, like graphene ribbons. Their inherent Raman and optic spectra should be individual for each 2HG-SL similar to single-walled carbon nanotubes [7].

We have also shown that the electronic density near the Fermi energy is localized between the 2H lines in the (3$n$,0) strips of 2HG-SL heterostructures. Hence, electronic waveguides and heterostructures and, therefore, new nanoelectronic devices based on them, can be obtained by creating lines of pairs of H-atoms adsorbed on graphene. The nanoprint method should allow the production of quantum electronic chips on a single graphite sheet, in an analogy with planar optics chips.

We are grateful to the Joint Supercomputer Center of the Russian Academy of Sciences for the possibility of using a cluster computer for quantum-chemical calculations, to I.V. Stankevich, V.A. Geyler and L. Biro for fruitful discussions. Molecular orbitals were visualized using the Molekel 4.3 program. The geometry of all presented structures was visualized by ChemCraft software (http://www.chemcraftprog.com). This work was supported by the Russian Foundation for Basic Research (project no. 05-02-17443) and Deutsche Forschungsgemein-schaft/Russian Academy of Sciences (DFG/RAS, project no. 436 RUS 113/785).